# Strong magnetophonon oscillations in extra-large graphene

P. Kumaravadivel[1,2], M. T. Greenaway[3,4], D. Perello[1,2], A. Berdyugin[1], J. Birkbeck[1,2], J. Wengraf[1,5], S. Liu[6], J. H. Edgar[6], A. K. Geim[1,2], L. Eaves[1,4], R. Krishna Kumar[1]

[1]School of Physics & Astronomy, University of Manchester, Manchester M13 9PL, UK

[2]National Graphene Institute, University of Manchester, Manchester M13 9PL, UK

[3]Department of Physics, Loughborough University, Loughborough LE11 3TU, UK

[4]School of Physics & Astronomy, University of Nottingham, Nottingham NG7 2RD, UK

[5]Department of Physics, University of Lancaster, Lancaster LA1 4YW, UK

[6]Department of Chemical Engineering, Kansas State University, Manhattan, KS 66506, USA

**Van der Waals materials and their heterostructures offer a versatile platform for studying a variety of quantum transport phenomena due to their unique crystalline properties and the unprecedented ability in tuning their electronic spectrum. However, most experiments are limited to devices that have lateral dimensions of only a few micrometres. Here, we perform magnetotransport measurements on graphene/hexagonal boron-nitride Hall bars and show that wider devices reveal additional quantum effects. In devices wider than ten micrometres we observe pronounced magnetophonon oscillations that are caused by resonant scattering of Landau-quantised Dirac electrons by acoustic phonons in graphene. The study allows us to accurately determine graphene's low energy phonon dispersion curves and shows that transverse acoustic modes cause most of the phonon scattering. Our work highlights the crucial importance of device width when probing quantum effects in van der Waals heterostructures and also demonstrates a precise, spectroscopic method for studying electron-phonon interactions in van der Waals heterostructures.**

Two-dimensional electronic systems exhibit a rich variety of quantum phenomena[1,2]. The advent of graphene has not only provided a way to study these phenomena in the quasi-relativistic spectrum, but has also extended their experimental range[3,4], made some observations much clearer[5,6,7,8] and, of course, revealed many new effects[9,10,11,12]. These advances are mostly due to graphene's intrinsically high carrier mobility that is preserved by state-of-the-art heterostructure engineering in which graphene is encapsulated between hexagonal-boron nitride layers[13,14] and electrically tuned with atomically smooth metallic gates[8,15]. Nonetheless, one of the first discoveries in quantum transport, well known for over fifty years[16,17], has remained conspicuously absent in graphene - magnetophonon oscillations[18,19].

In the presence of an applied magnetic field (*B*), electrons in pristine crystals become localised in closed orbits and their spectra take the form of quantised Landau levels (LLs) separated by energy gaps. However, an electrical current can still flow in the bulk due to carriers resonantly scattering between neighbouring orbits by the absorption or emission of phonons with energies equal to the LL spacing[19] (Fig. 1a). In a semi-classical model, the resonant transitions occur between orbits which just touch in real space and induce "figure of eight" trajectories[20] (Fig. 1b), corresponding quantum mechanically to strong overlap of the tails of their wave functions in the vicinity of their classical turning point. This effect, known as magnetophonon resonance (MPR) causes magnetoresistance oscillations that are periodic in inverse magnetic field[19,21,]. Whereas magnetophonon oscillations have been used extensively for studying carrier-phonon interactions in bulk Si and Ge[22], semiconducting alloys[18] and heterostructures[23,24,25], there have been no reported observations in any van der Waals crystal, not even graphene, despite its exceptional electronic quality.

In this article, we consider a subtle yet crucial aspect concerning the design of electronic devices based on graphene, namely the lateral size of the conducting channel. It has so far remained small, only a few micrometres in most quantum transport experiments. Our measurements using graphene Hall bars of different widths show that wider samples start exhibiting pronounced magnetophonon oscillations.

**Phonon scattering in wide graphene channels**

Our experiments involved magnetotransport measurements on graphene Hall bars encapsulated by hexagonal-boron nitride, with particular attention paid to 'wide' devices with channel widths $W > 10$ μm. An optical image of one of our widest devices is shown in Fig. 1c (See Supplementary Section 1 for details of device fabrication). Because the electron-phonon coupling is so weak in graphene[26], charge carriers scatter more frequently at the device edges of micron-sized samples rather than with phonons in the bulk, especially at low temperature[27] ($T$). This is evident when comparing the Drude mean free path ($L_{MFP}$) for devices of different $W$ and a fixed carrier density ($n$) of holes (Fig. 1d). At 5 K, all devices exhibit size-limited mobility ($L_{MFP} \geq W$) because carriers propagate ballistically until they collide with the edges of the conducting channel. Even at 50 K, scattering is still dominated by the edges in most of our devices and $L_{MFP}$ increases linearly with $W$. However, at these higher temperatures we find that $L_{MFP}$ saturates around 8 μm (green line in Fig. 1d) and does not increase upon further widening of the device channel. This saturating behaviour tells us that $L_{MFP}$ is no longer dependent on the device width and carriers scatter mostly with phonons in the bulk ($L_{MFP} < W$). In effect, widening the channel makes our measurement more sensitive to bulk phenomena rather than edge effects.

**Width dependent magnetophonon oscillations**

The main observation of our work is presented in Figure. 1e, which plots the longitudinal resistance ($R_{xx}$) of a 15 μm wide Hall bar (Fig. 1c) as a function of $B$, at two $T$ and fixed $n$. At 5 K we observe two distinct oscillatory features. The first, at relatively low $B < 0.2$ T, are the well-established semi-classical geometrical resonances that occur due to magnetic focussing of carriers between current and voltage probes[4] (Supplementary Section 2). At higher $B$ (~ 1 T), quantised cyclotron orbits are formed and we observe 1/$B$-periodic Shubnikov de Haas (SdH) oscillations. Their origin is confirmed by noting that the charge carrier density $n = 4e/(h\Delta(B^{-1}))$ extracted from the SdH period ($\Delta(B^{-1})$) agrees with that determined by Hall effect measurements (Fig. 1e inset). At 50 K, the low-field geometric oscillations remain visible although their relative amplitudes are suppressed due to the reduced carrier mean free path. However, at higher $|B| > 0.2$ T, a new set of oscillations appears with five clear maxima (indicated by red arrows in Fig. 1e). These high-$T$ oscillations are also periodic in $1/B$ but are distinguished by their markedly slower period. In contrast to $R_{xx}$, the Hall resistance, $R_{xy}$, shows no oscillatory features and has the same value at both $T$ (Fig. 1e inset), confirming that $n$ does not change upon warming the sample.

The observation of the high-$T$ oscillations depends critically on the sample width. This is shown in Fig. 1f which plots the normalised magnetoresistance, $R_{xx}/R_{xx\ (B\ =\ 0\ T)}$, for devices with different $W$ at fixed $T$ and $n$. We note that the bulk channels in all our devices are intrinsically clean and free from defects (probed by ballistic transport experiments in Supplementary Section 3). Nonetheless, whereas these oscillations are well developed in the widest devices (resonances marked by purple arrows), they are poorly resolved for devices with $W < 8$ μm and completely absent in the narrowest one ($W = 1.5$ μm). As described below, we identify these size dependent, high-$T$ oscillations with MPR. '

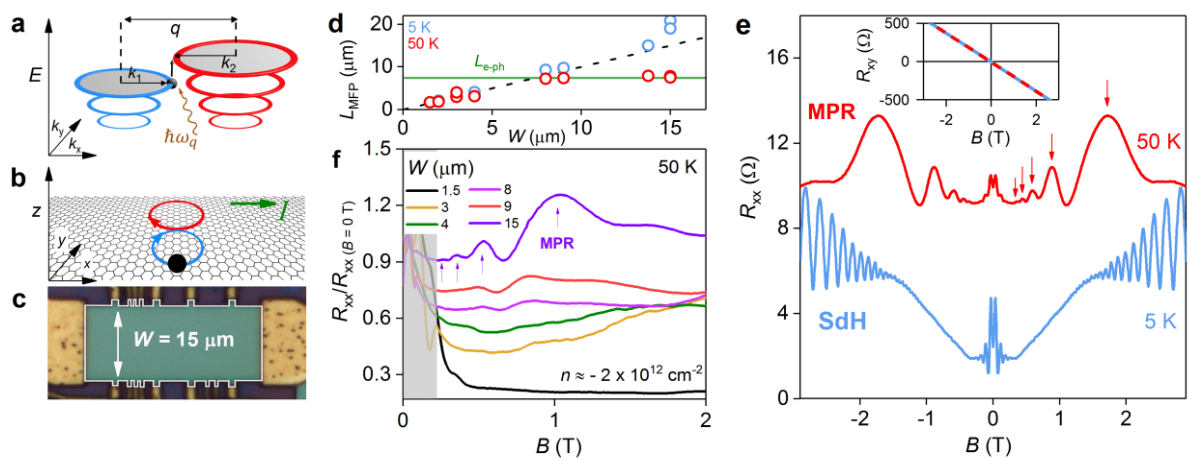

**Figure 1| Size dependent magnetoresistance oscillations in mesoscopic graphene devices**. **a**, Landau level spectra of graphene. The diagram illustrates a carrier with momentum $k_1$ (black sphere) making a transition between Landau levels (blue and red rings) by resonant absorption of a phonon (brown arrow) with momentum $q = |k_1 - k_2|$ and energy $\hbar\omega_q$. Solid black arrows represent the magnitudes of wavevectors $k_1$, $k_2$ and $q$. **b**, The semiclassical motion of a carrier (black sphere) in real space for the resonance condition sketched in **a**. The red and blue circles which touch at a "tangent" point represent the initial and final semiclassical cyclotron orbits of two Landau quantised states between which an electron can be scattered resonantly by a phonon. During resonant scattering, the carrier motion follows a path that resembles the number 8 which we refer to as a "figure of eight" trajectory. The red and blue arrows show the motion of the charge carriers along this trajectory. The green arrow indicates direction of the applied current $I$ **c**, optical image of a graphene device with $W = 15$ μm. The edges of the mesa are indicated by the white solid line. **d**, Open circles plot experimentally determined Drude mean free path $L_{MFP}$ as a function of $W$ for two $T$ and fixed $n = -2 \times 10^{12}$ cm$^{-2}$. Black dashed line traces points on the axis where $L_{MFP} = W$. Solid green line marks the phonon-limited mean free path ($L_{e\text{-}ph}$) at 50 K. Our measurements focussed on the valence band because our wide devices exhibited higher electronic quality for hole doping **e**, longitudinal magnetoresistance data $R_{xx}(B)$ for fixed $n = -3.3 \times 10^{12}$ cm$^{-2}$ measured in our wide device (**c**) at two different $T$. The 50 K curve is off-set vertically for clarity. Inset: Hall resistance $R_{xy}(B)$ measured simultaneously as $R_{xx}$. The solid blue and dashed red lines are data measured at 5 and 50 K respectively. **f**, $R_{xx}/R_{xx\ (B\ =\ 0\ T)}$ measured at fixed $n$ and $T$ in several devices of different $W$. The shaded area close to $B = 0$ contains semi-classical effects[4].

A defining feature of magnetophonon oscillations is their unique non-monotonic temperature dependence, in which their amplitude first increases with $T$ and then decays[25]. Fig. 2a shows the temperature dependence of $R_{xx}(B)$ for fixed $n$ between 5 and 100 K (5 K steps) for another wide Hall bar device ($W$ = 15 μm). In this sample, weak magnetophonon oscillations already appear at 5 K in the field range between the geometric and the SdH oscillations. The resonances are labelled $p$ = 1 to 5, where the integer $p$ refers to the number of LL spacings that are crossed during the transition; $p$ = 1 corresponds to scattering between LLs adjacent in energy (Fig. 1a). With increasing $T$, the magnetophonon oscillations become more pronounced as more phonons are thermally activated, whilst the SdH oscillations are strongly suppressed. Although both phenomena require carriers that exhibit coherent cyclotron orbits ($\mu B$ > 1, where μ is the carrier mobility), MPR is not obscured by smearing of the Fermi–Dirac distribution across Landau gaps[25]; rather it is enhanced due to an increased number of unoccupied states into which carriers can scatter. Hence magnetophonon oscillations persist to higher $T$ than SdH oscillations. However, they are eventually damped at high enough $T$ (Fig. 2b) when LLs become broadened by additional scattering ($\mu B \sim 1$). This non-monotonic behaviour is better visualised in Fig. 2c which plots the oscillatory amplitudes ($\Delta R_{xx}$) as a function of $T$. Notably, the amplitude of all resonances peak at $T$ below 60 K, corresponding to a thermal energy of a few meV.

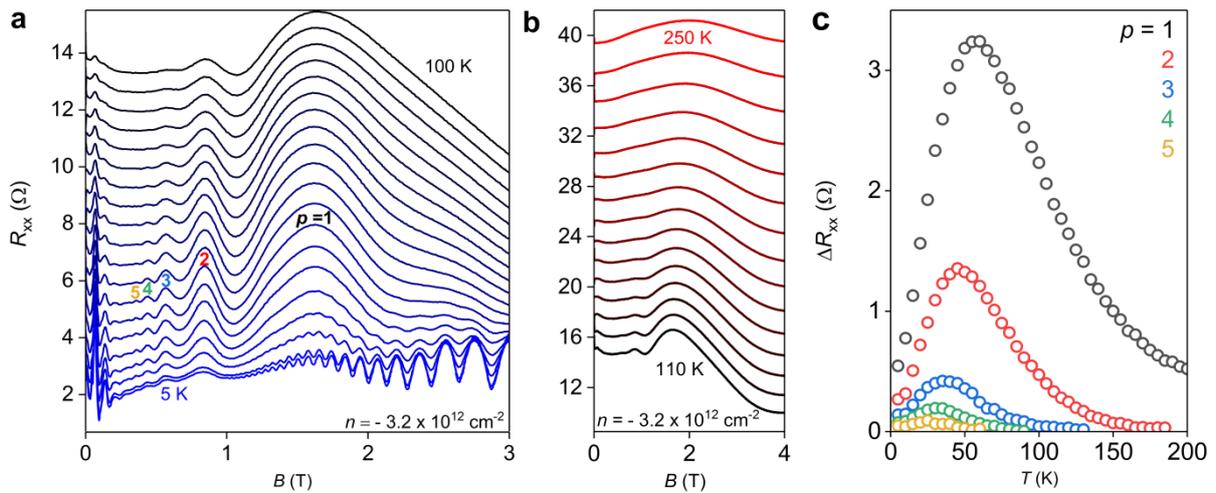

**Figure 2| Temperature dependence of the magnetophonon effect. a**, Magnetoresistance $R_{xx}(B)$ for $T$ between 5 K (blue curve) and 100 K (black curve) in 5 K steps for fixed $n$ measured in another Hall bar with $W$ = 15 μm. **b**, Extended data set of **a** showing high $T$ behaviour (10 K steps). **c,** Temperature dependence of the amplitude of MPR peaks, $\Delta R_{xx}(T)$, indicated in **a** by colour coded letters, $p$ = 1 - 5.

**MPR spectroscopy**

For the doping levels and $B$-fields at which the oscillations occur, the charge carriers occupy high-index LLs ($N \sim 20$ for $p$ = 1) separated by small energy gaps ($\sim$ 5 meV) with a classical cyclotron radius up to $R_c \sim \hbar k_F/eB \sim 300$ nm, where $k_F$ is the Fermi-wave vector. Resonant inter-LL transitions occur due to inelastic scattering by low-energy acoustic phonons that induce figure of eight trajectories (Fig. 1b). This type of trajectory occurs with high probability because the wavefunctions of the initial (blue circle in Fig. 1b) and final states (red circle) have a large spatial overlap where they touch in real space[24]. During figure of eight trajectories, the velocity of the carrier is reversed at the intersection of the two cyclotron orbits (see arrows in Fig. 1b). This process requires a phonon of specific momentum $q \approx 2k_F \sim 10^9$ m$^{-1}$ and energy $\hbar\omega_q(2k_F) \sim 5$ meV that can back-scatter the carriers during the inter-LL transition. Energy and momentum conservation for such a process requires that $E_{N+p} - E_N = \hbar\omega_q(2k_F)$, where $E_N$ is the energy of an electron in the $N^{th}$ LL, so that resonances occur at $B$ values given by

$$B_p = \frac{nhv_s}{pev_F} \tag{1}$$

(See Supplementary Section 4 for a detailed derivation). Here, $v_F$ and $v_s$ are the Fermi-velocity and low-energy acoustic phonon velocity in graphene respectively. This resonant condition is unique to massless Dirac electrons and is strikingly different to the case of massive electrons in a conventional two dimensional electron gas (2DEG)[24] system where $B_p$ scales with $n^{0.5}$. On resonance, inelastic scattering between neighbouring orbits (Fig. 1b) gives rise to a finite and dissipative current in the bulk. This behaviour causes maxima in $\rho_{xx}$ at $B_p$; the 1/B periodicity results in oscillations described by $\Delta\rho_{xx} \sim e^{-\gamma/B}\cos(2\pi B_F/B)$ where $B_F \equiv pB_p$ and the factor $\gamma$ depends on temperature[28]. Equation (1) predicts that the position of maxima scales linearly with $n$. With this in mind, Figs. 3a,b plot maps of $R_{xx}(n, B)$ for one of our 15 μm devices at 5 K (Fig. 3a) and 50 K (Fig. 3b). In addition to the typical SdH Landau fan structure that is dominant at low $T$ (filling factors, ν, are marked by blue dashed lines), the maps reveal a broader set of fans at lower $B$ that are more prominent at 50 K (Fig. 3b). They are caused by MPR ($p$ values are labelled in red) and demonstrate that their frequency scales linearly with $n$. Furthermore, the positions of MPR peaks in Fig. 3b can be fitted precisely by equation (1) (red dashed lines) with a constant $v_s/v_F = 0.0128$. By studying the temperature dependence of SdH oscillations in our graphene devices (Supplementary Section 5), we extract $v_F$ and determine $v_s$ accordingly. We note that $v_F$ showed no significant changes for different $n$, as expected for graphene devices on dielectric substrates at high doping[29] because e-e interactions that cause velocity renormalization[30] are heavily screened. Hence, using the extracted $v_F = 1.06 \pm 0.05 \times 10^6$ ms$^{-1}$, we determined the phonon velocity as $v_s = 13.6 \pm 0.7$ kms$^{-1}$. This value is close to the speed of transverse acoustic (TA) phonons in graphene (~ 13 kms$^1$) calculated in numerous theoretical works[31,32,33,34,35]. Therefore, we infer our oscillations arise from inter-LL scattering by low energy and linearly dispersed TA phonons.

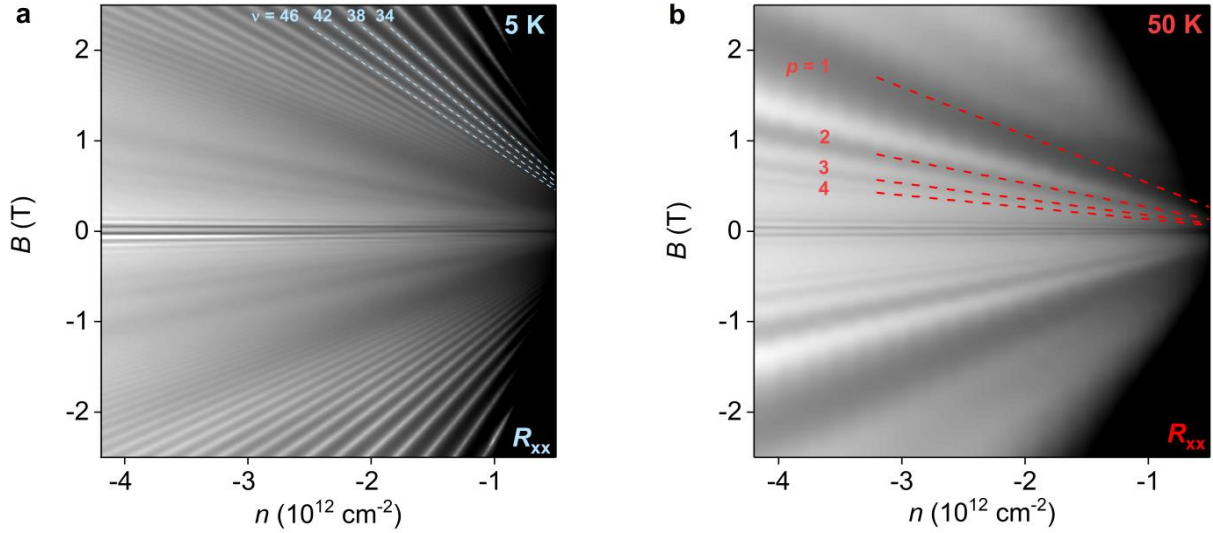

**Figure 3| Density dependence of magnetophonon oscillations. a,** longitudinal resistance $R_{xx}$ (grey scale map) as a function of $n$ and $B$ measured at 5 K ($W$ = 15 μm). Logarithmic grey scale; White: 1 Ω to black: 15 Ω. The blue dashed lines trace Landau gaps corresponding to high filling factors ν = $nh/Be$. **b,** same as **a** measured at 50 K. Logarithmic grey scale; White: 5.5 Ω to black: 18 Ω. The red dashed lies plot equation (1) for $p$ = 1 to 4 which corresponds to charge carriers scattering with transitions across 1 to 4 Landau level spacings. . Features appearing for $B$ < 0.2 T are the semi-classical geometrical oscillations[4] not relevant in this work (see Supplementary Section 2 for details).

Equation (1) is generic for linearly dispersed phonons in graphene. This motivated us to search for MPR arising from longitudinal acoustic (LA) phonons, which should occur at higher $B$ due to their significantly higher $v_s$[34,35]. Careful inspection of the data in Fig. 1-2 shows that the $p$ = 1 resonance for TA phonons is followed by a weak shoulder-like feature at higher $B$. We therefore studied a dual-gated graphene device that permitted measurements at higher $n$ ~ -1 x $10^{13}$ cm$^{-2}$ which, according to equation (1), should better separate this feature from the TA resonances. Fig. 4a plots $R_{xx}(B)$ for this device for several $n$. Measurements at these high $n$ reveal the shoulder-like feature developing into a well-defined peak (indicated by coloured arrows). Its position ($B_{p=1}$) is accurately described by equation (1) with a constant $v_s/v_F$ = 0.0198. Using our extracted values of $v_F$, we obtained $v_s$ = 21.0 ± 1.0 kms$^{-1}$. This value is indeed close to that calculated for LA phonons in graphene[31,34,35], and hence we attribute this feature to inter-LL scattering by LA phonons.

Further validation of our model is presented in Fig. 4b, which plots the magnetophonon oscillation frequency for TA phonons ($B_F \equiv pB_p$) as a function of $n$ for several different devices (red symbols). It shows that a linear dependence (red line) fits the data to equation (1) for all our measured devices over a range of $n$ spanning an order of magnitude. The weaker LA resonance was also found to occur at the same $B_{p=1} = B_F$ in different devices (blue symbols). Furthermore, the data in Fig. 4b can be transformed directly into phonon dispersion curves (Inset of Fig. 4b) by noting that $q \approx 2k_F = 2(n\pi)^{0.5}$ and $\hbar\omega_q = (2eB_Fv_sv_F\hbar)^{0.5}$. The extended tunability of the carrier density in our dual-gated devices allows measurement of phonon branches up to wave vectors > $10^9$ m$^{-1}$. Note that these dispersion plots are significantly more precise than those measured by X-ray scattering experiments in graphite[36] (purple stars). Studies of magnetophonon oscillations thus enable an all-electrical measurement of the intrinsic phonon dispersion curves in gate-tuneable materials.

## Discussion

To understand why magnetophonon oscillations are absent in narrow samples, we first note that figure of eight trajectories (Fig. 1b) have a spatial extent ~ $4R_c$, which can reach values of several microns for the high-order resonances ($p > 3$). If the sample is too narrow, so that $4R_c$ is comparable to $W$, the carrier trajectories are skewed by elastic scattering at the device edges. In this case, they propagate along the edges of the device in skipping orbits[2], effectively "short-circuiting" the resistive behaviour of the bulk caused by MPR. However, if $W > 4R_c$, both MPR and skipping orbits contribute to $R_{xx}$. We can estimate the width of the device required to observe MPR by comparing the relative contributions of these two processes. Carriers that diffuse in MPR-induced figure of eight trajectories move a distance $2R_c$ in a characteristic time, $\tau_{e\text{-ph}} = L_{e\text{-ph}}/v_F$ with a drift velocity $v_{MPR} = 2R_c/\tau_{eph}$. This is significantly slower than skipping orbits which can have speeds approaching $v_F$. On the other hand, skipping orbits occupy only a width ~ $R_c$ at each edge, whereas MPR occurs approximately over the full width, $W$, of the bulk. By comparing these two contributions, we deduce that MPR dominates when $Wv_{MPR} \gtrsim 2R_c v_F$. This corresponds to the condition $W \gtrsim L_{e\text{-ph}}$, in good agreement with the measured data in Figs. 1d, f.

Our measurements provide an important insight into the intrinsic electron-phonon interaction in graphene: namely, the dominance of carrier scattering by low-energy TA phonons. This is in agreement with several theoretical works[35,37,38] and contrasts with a widely held view that deformation potential scattering by LA phonons prevails over TA phonons[39]. To investigate this point further, we calculated magnetoresistance using the Kubo formula[40] (Supplementary Section 6). A typical calculation is shown in Fig. 4c, which plots the contribution ($\Delta\rho_{xx}$) of MPR for TA and LA phonon velocities of $v_s$ = 13.6 and 21.4 kms$^{-1}$, respectively[35] and the Fermi velocity[41] $v_F = 1 \times 10^6$ ms$^{-1}$. It accurately describes the oscillatory form of the measured data. Such good agreement is only possible when our calculations include the effect of carrier screening[35,38,42,43] which significantly reduces the electron-LA phonon deformation potential coupling. Without screening, LA phonons would dominate the observed MPR (Supplementary Section 6). Our results therefore highlight the importance of carrier screening on electron-phonon interactions and thus helps resolve a long-standing discussion of the relative importance of LA[44,39] and TA[37,38,43] phonon scattering in graphene.

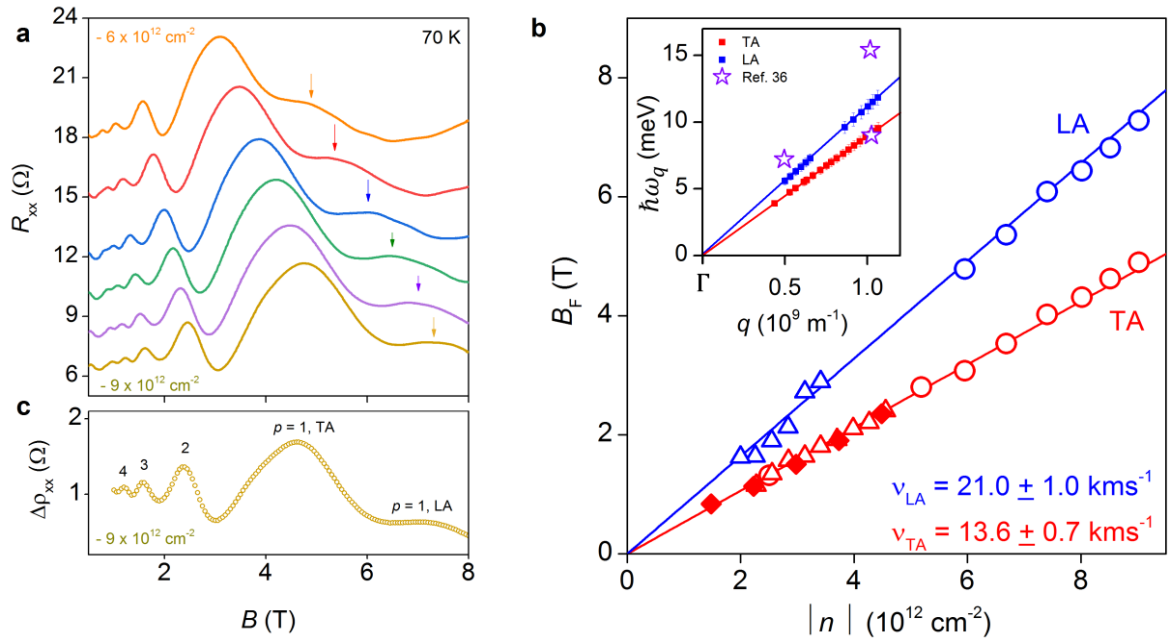

**Figure 4 | Phonon spectroscopy in graphene by measurement of magnetophonon oscillations. a,** Longitudinal resistance $R_{xx}$ as a function of $B$ measured for several high $n$ of holes in our wide ($W$ = 13.8 μm) dual-gated graphene Hall bar. The curves have been off-set vertically for clarity. **b,** Red symbols plot the fundamental frequency $B_F \equiv pB_p$ of magnetophonon oscillations as a function of absolute $n$ for three different devices; open circles correspond to the dual-gated device which allowed high doping. The blue symbols mark the positions $B_{p=1}$ of the broad peak which appears clearly at high $n$ (indicated by coloured arrows in **a**). The red and blue solid lines represents equation (1) with $v_s/v_F$ = 0.0128 and 0.0198 respectively. Knowing $v_F$ (Supplementary section 5) we extract the TA ($v_{TA}$) and LA ($v_{LA}$) phonon velocities. Inset: Data in main panel transformed to phonon dispersion curves. Coloured squares – experimental data points (error bars reflect the error in the experimentally extracted $v_F$), solid lines plot the equation $\hbar\omega_q = \hbar v_s q$ (same $v_s$ as in main panel), purple stars – data taken from ref. 36. **c,** calculation of the oscillatory part of the resistivity $\Delta\rho_{xx}(\Omega)$ using the Kubo formula (see Supplementary Section 6 for details).

**Conclusion**

To conclude, we report the first observation of magnetophonon oscillations in graphene, where the Dirac spectrum strongly modifies the resonant condition compared to previously studied electronic systems. Other two-dimensional crystals can also be expected to exhibit this phenomenon. The oscillations enable the study of low-energy acoustic-phonon modes that are generally inaccessible by Raman spectroscopy[45,46]. Our measurements combined with the Kubo calculations provide strong evidence that TA phonons limit temperature-dependent mobility in graphene[35,37,38]. Most importantly, graphene's transport properties are shown to strongly depend on device size, even for conducting channels as wide as several microns. This should motivate further experiments on graphene and related two-dimensional materials in a macroscopic regime beyond the scope of previous mesoscopic devices.

## Methods

For measuring resistance in our graphene devices, we used standard low-frequency AC measurement techniques with a lock-in amplifier at 10–30 Hz. The measurements of $R_{xx}(\Omega) = V_{xx}/I_{xx}$ are obtained by driving a small AC excitation current ($I_{xx}$ = 0.1–1 µA) down the length of the Hall bar while simultaneously measuring the four probe voltage drop ($V_{xx}$) between two side contacts located on the edge of the Hall bar devices (Fig. 1c). We tune the Fermi level in our graphene devices by applying a DC voltage between the silicon substrate and the graphene channel, where the $SiO_2$ and bottom hexagonal boron nitride encapsulation layer serve as the dielectric (see Supplementary section 1 for details on device fabrication). In our top gated device (see Supplementary Fig. 1), we simultaneously apply a potential to the metal top gate which allowed us to reach higher doping levels (see Fig. 4). All measurements were performed inside a variable temperature inset of a wet helium-4 flow cryostat that allowed us to carry out temperature-dependent magnetotransport measurements using a cold superconducting magnet.

## Supplementary Information

### 1 Device Fabrication

The hexagonal boron nitride (hBN) encapsulated graphene heterostructures were assembled using a dry-peel transfer method. Graphene and hBN flakes were first exfoliated onto $O_2$/Ar plasma cleaned $SiO_2$/Si substrates. The appropriate flakes were then identified by long exposure dark field optical imaging. For the hBN encapsulation layers, we used flakes that were ~ 25 – 100 nm thick. The selected flakes were then assembled using a polypropyl carbonate (PPC) coated Polydimethylsiloxane (PDMS) stamp placed on a glass slide attached to a high precision XYZ micromanipulator[1]. First, the top hBN encapsulation layer was picked up using the PPC/PDMS stamp. This was done at a fixed substrate temperature ~ $50^0$ C. The hBN flake attached to the stamp was then used to pick up the graphene flake. During this process, the substrate temperature was fixed at $65^o$ C whilst smoothly contacting hBN to graphene by fine Z adjustments of the micromanipulator. Once fully contacted, at $50^o$ C, the resultant hBN/graphene stack was peeled off from the substrate onto the PPC/PDMS stamp. The stack was then placed down on the bottom hBN flake (substrate temperature at $65^o$ C) to fully encapsulate the graphene layer.

Once the heterostructure was prepared, we performed standard electron beam lithography techniques to create the Hall bar geometry. First, we patterned a polymethyl methacrylate (PMMA) mask on the stack to define contact regions leading up to the device channel. The regions unprotected by the mask were etched away using $CHF_3$ + $O_2$ reactive ion etching (RIE), forming narrow trenches. Metal contacts (5 nm Cr/70 nm Au) were then evaporated into the trenches which form high-quality contacts to the graphene edge[2,3]. Next, the same lithography and RIE etching procedures were used to pattern the Hall bar mesa. For one of our devices, we also patterned a metallic top gate above the heterostructure (Fig S1) which allowed us to achieve higher doping levels.

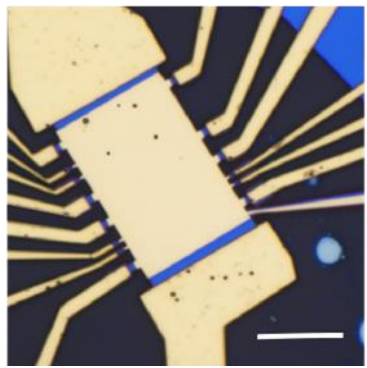

**Figure S1| Wide graphene Hall bars.** Optical image of our wide top-gated device ($W$ = 13.8 μm). White scale bar is 10 μm.

## 2 Magnetic Focussing in wide graphene Hall bars

At low temperature, the charge carriers in our graphene/hBN devices propagate ballistically for several micrometres before scattering. For certain measurement geometries, the 4-probe resistance is governed by direct transmission of charge carriers between current and voltage probes. An example is illustrated in Figure S2a which describes the measurement scheme for a transverse magnetic focussing experiment. In the presence of a perpendicular magnetic field ($B$), ballistic carriers injected from the current source follow curved trajectories and the measured 4-probe resistance ($R_{TMF}$) exhibits maxima for particular values of $B$ when the carriers are focussed directly into the collector voltage probe (coloured arrows in Fig. S2a). This occurs when the cyclotron radius ($R_c \sim \hbar k_F/eB$) becomes commensurate with the distance ($L$) between the current injector and voltage collector. The resonance condition is given by

$$B = \frac{2\hbar k_F s}{eL} \quad \text{(S1)}$$

where $k_F$ is the Fermi wave-number and ($s$-1) is an integer number that describes the number of reflections at the device edge. For example, $s = 2$ describes the trajectories traced by the red arrows in Fig. S2a. Figure. S2b shows $R_{TMF}$ ($B$) at fixed $n$ measured in the geometry illustrated in Fig. S2a. For negative values of $B$, we find a set of maxima that are equally spaced with a period $\Delta B = 2\hbar k_F/eL$. No resonances are observed for positive $B$ because the Lorentz force acts in the opposite direction and bends trajectories away from the voltage collector. Figure. S2c plots maps of $R_{TMF}$ ($B$,$n$) for different hole doping. We find the resonances are shifted to higher $B$ for higher $n$, in agreement with Eq. (S1) and the fact that the cyclotron radius is larger for higher $n$. Notably, all the resonances can be described by Eq. (S1) with $L = 7.4$ μm (dashed lines in Fig. S2c). This value corresponds to the distance between the current injector and voltage collector (labelled in Fig. S2a) and validates the dependence of the oscillatory features given by Eq. (S1).

Magnetic focussing resonances can also appear in a standard longitudinal resistance measurement ($R_{xx}$) if the voltage probes are located too close to the current contacts (see main text). This occurs if they are closer than the width of the channel and is typically the case in our wider samples (Fig. S2d) because of the limited size of exfoliated flakes and clean areas available for patterning devices[4,5]. Figure. S2e plots $R_{xx}$ ($B$) for a fixed $n$ and $T$ in non-quantizing $B$ fields. We find a set of resonances that are periodic in $B$ and occur at higher $B$ for higher $n$ (Fig. S2f). This behaviour resembles that of magnetic focussing although these curves are distinct from a typical measurement (Fig. S2a-c). First, the current injectors are rather wide (15 μm). Second, the oscillations are slightly phase shifted; whereas resonances are equally spaced with a period $\Delta B$, the first (labelled 1 in Fig. S2e) occurs at a value approximately $\Delta B/2$. To understand the origin of those resonances, we used Eq. (S1) to extract $L$ from the oscillation period $\Delta B = 2\hbar k_F/eL$. We carried out this analysis for several different doping levels (Fig. S2f) and found approximately the same $L \approx 5.6$ μm. This corresponds roughly to the distance between current and voltage probes and suggests that the dominant contribution to the measurement originates from magnetic focussing where carriers are injected from the corners of the device (yellow arrows in Fig. S2d). Further work is required to understand the details of these magnetic focussing resonances.

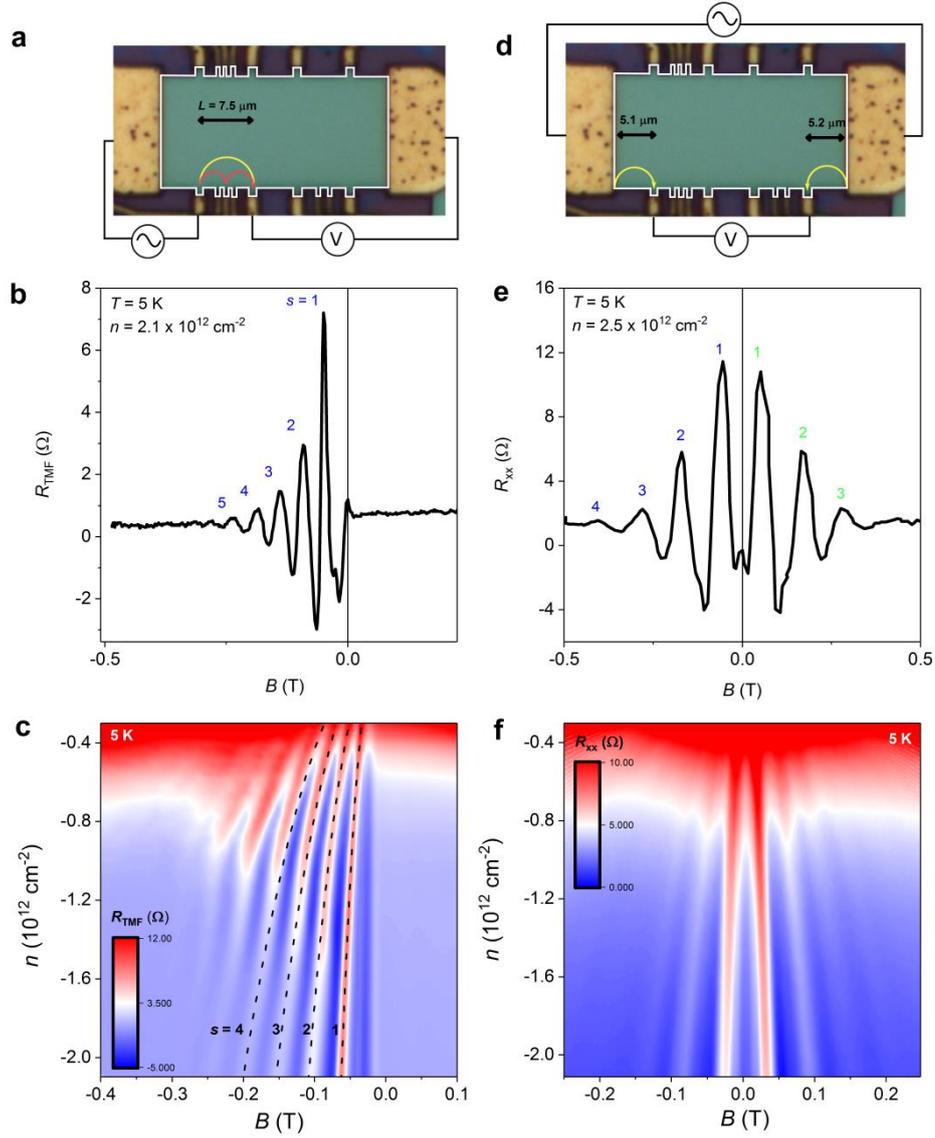

**Figure S2| a,** Measurement scheme for a typical transverse magnetic focussing experiment performed in a graphene device with $W$ = 15 um. Yellow and red lines trace trajectories of electrons under the resonance condition (1) for $s$ = 1 and 2 respectively. **b,** $R_{TMF}$ ($B$) for fixed $n$ and $T$. **c,** Magnetic focussing maps, $R_{TMF}$ ($B$, $n$), for the configuration specified in **a**. Dashed lines are fits to Eq. (S1) with $L$ = 7.4 μm and different $s$. **d** Measurement scheme for $R_{xx}$ geometry; the yellow arrows depict trajectories of charge carriers in both negative and positive $B$-fields. **e,** $R_{xx}$ ($B$) for a fixed $n$ and $T$. **f,** maps $R_{xx}$ ($B$, $n$).

### 3 Quasi-ballistic device channels

Figure. 1f of the main text shows that magnetophonon oscillations in graphene appear only in samples in which the channel has a sufficiently large width ($W$). We argue that the size is the important variable because all our devices exhibit similar electronic quality in the bulk; their conducting channels are relatively free from defects/impurities so that low-temperature mobility is limited only by scattering at the edges[6]. To prove this, we performed magnetic focussing experiments (described in Supplementary Section 2) on all our devices. The observation of magnetic focussing resonances requires carriers that propagate ballistically without scattering, thus proving there are no scattering centres along the path between the current injector and voltage collector. Here we present data taken from transverse magnetic focussing experiments performed in our narrowest ($W$ = 1.5 μm) and widest ($W$ = 15 μm) samples. The measurement geometries for each device are sketched in Fig. S3a-b. Figure S3c plots $R_{TMF}$ ($n,B$) measured in the narrowest device in which injector and collector probes are separated by $L$ = 1.5 μm. We find pronounced magnetic focussing resonances up to the fourth order ($s$ = 4). We note that similar resonances also appear between any pairs of contacts located in different regions of the device, thus providing further confirmation of the quality and ballistic nature of our channels. Even in our widest samples (Fig. S3b) we detect ballistic electrons focused at voltage probes 20 μm away from current injector (Fig. S3d).

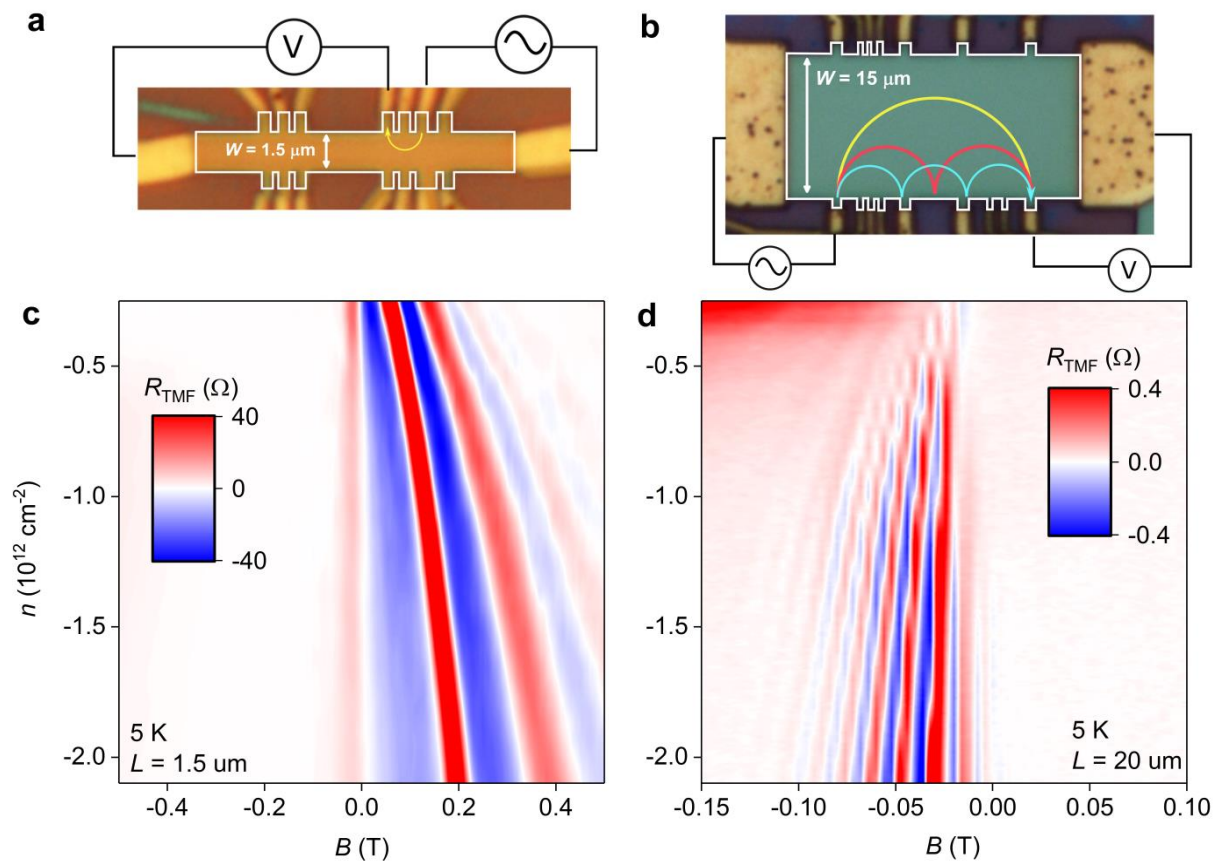

**Figure S3|** **a,b,** Measurement scheme for a magnetic focussing experiment performed in our narrowest (**a**) and widest (**b**) Hall bar devices (the mesa are contoured in white). The yellow, red and blue curved arrows indicate electron trajectories corresponding to resonances predicted by Eq (S1) with $s$ = 1, 2 and 3 respectively. **c,d** Transverse magnetic focussing maps, $R_{TMF}$ ($n, B$) measured at 5 K in the configurations shown in **a** and **b** respectively.

## 4 Semiclassical model of magnetophonon resonance in monolayer graphene

The magnetoresistivity of the device is given by the approximation

$$\rho_{yy} \approx \frac{\sigma_{xx}}{\sigma_{xy}^2} \approx \rho_{xx}, \qquad (S2)$$

since the Hall component of the conductance tensor, $\sigma_{xy} \gg \sigma_{xx}$ for these experimental conditions. The longitudinal conductance tensor component, $\sigma_{xx}$, is determined by the rate of drift of a carrier's cyclotron orbit centre. This process is illustrated semiclassically in Fig. 1b of the main text which shows the shift caused by an inelastic scattering-induced figure–of-eight transition in *k*-space. The energy absorbed (or emitted) by the carrier undergoing an inter-Landau level (LL) transition is given by the energy difference between its initial (*N*) and final (*N* ± *p*) states, where *N* is the LL index and *p* is a positive integer. The quantised energy spectrum of monolayer graphene is given by

$$E_N = \mathrm{sgn}(N)\sqrt{2|N|}\frac{\hbar v_F}{l_B}, \qquad (S3)$$

where $v_F$ is the Fermi velocity of graphene and $l_B = \sqrt{\hbar/eB}$ is the quantum magnetic length. For a figure-of-8 transition, an electron in a LL with index $N$ and orbit radius in *k*-space given by $\kappa_c = \sqrt{2N}/l_B$ makes an inelastic transition to a level with index $N' = N \pm p$ by absorbing or emitting a phonon with wave vector, $q$, so that its final radius $\kappa'_c = q - \kappa_c$. In real space, the corresponding classical orbits have radii $R_c = l_B^2 \kappa_c$. The wavevector, $q$ induces a shift in real space of the orbit centre $\Delta X = l_B^2 q$ and hence provides a contribution to the current. An excellent fit to the measured period of the magnetophonon oscillations is obtained by considering scattering by linearly dispersed acoustic phonons with energy $\hbar \omega = \hbar v_s q = \hbar v_s (\kappa'_c + \kappa_c)$, where $v_s$ refers to velocity of either longitudinal acoustic (LA) or transverse acoustic (TA) phonons. The resonant conditions for absorption or emission processes are given by

$$\hbar\omega = \frac{\hbar v_s}{l_B}\left(\sqrt{2(N \pm p)} + \sqrt{2N}\right) = \pm \frac{\hbar v_F}{l_B}\left(\sqrt{2(N \pm p)} - \sqrt{2N}\right), \qquad (S4)$$

where *p* is the change in LL index. This can be expressed in the following form

$$N = \frac{pv_s}{4v_F}\left(\frac{v_F}{v_s} - 1\right)^2 \approx \frac{pv_F}{4v_s}. \qquad (S5)$$

The approximation holds since $v_s \ll v_F$. Only electrons within the energy range $\sim(E_F \pm 2k_B T)$, can scatter between filled and empty states by emitting or absorbing a phonon. The Fermi energy $E_F \gg 2k_B T$ over the temperature range of our measurements. The LL index, *N,* of the carriers undergoing MPR transitions around the Fermi energy is proportional to the carrier density, $n = 4eBN/h$, where the factor 4 corresponds to the two-fold valley and spin degeneracies. Combining this relation with (S5) we obtain the magnetic field position $B_p$ of the $p^{th}$ resonant peak:

$$B_p = \frac{nhv_F}{pev_s}\left(\frac{v_F}{v_s} - 1\right)^{-2} \approx \frac{nhv_s}{pev_F}. \qquad (S6)$$

Therefore, the magnetophonon resonance oscillations are periodic in 1/*B* with a frequency $B_F = pB_p$, as discussed in the main text.

## 5 Fermi velocity of charge carriers in graphene encapsulated with hexagonal boron nitride

Figure. 3 of the main text demonstrates that the position of maxima in the resistance caused by magnetophonon resonance can be described by equation (S6) with the only fitting parameter being the ratio of the phonon speed to the Fermi velocity in graphene $v_s/v_F$. Knowing the Fermi velocity of graphene we are able to measure the speed of the phonons responsible for the observed effect. It can be obtained by studying temperature dependence of Shubnikov de Haas (SdH) oscillations[7] and its measurement allows us to extract $v_s$ without any fitting parameters. Figure S4a plots the temperature dependence of the magnetoresistance $\rho_{xx}(B)$ for one of our graphene devices encapsulated with hBN. We observe pronounced 1/B-periodic SdH oscillations that are almost completely damped at 50 K (red curve in Fig. S4a). The inset of Figure S4a plots the amplitude of one of the oscillations (indicated by black arrow in Fig S4a) as a function of $T$. It shows that the data can be fitted precisely by the Lifshitz-Kosevich formula (solid red line) which allows us to extract the effective mass $m^*$. We repeated such measurement and analysis of SdH oscillations in different devices and for a range of n between 1 and 4 x $10^{12}$ cm$^{-2}$. These density dependent measurements of $m^*$ reveal a linear dependence of $m^*$ as function of the Fermi wavevector $k_F = (n\pi)^{0.5}$ (Fig. S4b), as previously observed for graphene on dielectric substrates[7,8]. Fitting the experimental data with the standard equation for $m^* = \hbar k_F/v_F$ in graphene[9] allows us to extract $v_F = 1.06 \pm 0.05 \times 10^6$ ms$^{-1}$. This value is in good agreement with previous measurements[7,8] and that which is typically expected for graphene[9] at these relatively high $n$ where velocity renormalisation due to e-e interaction[10] is negligible.

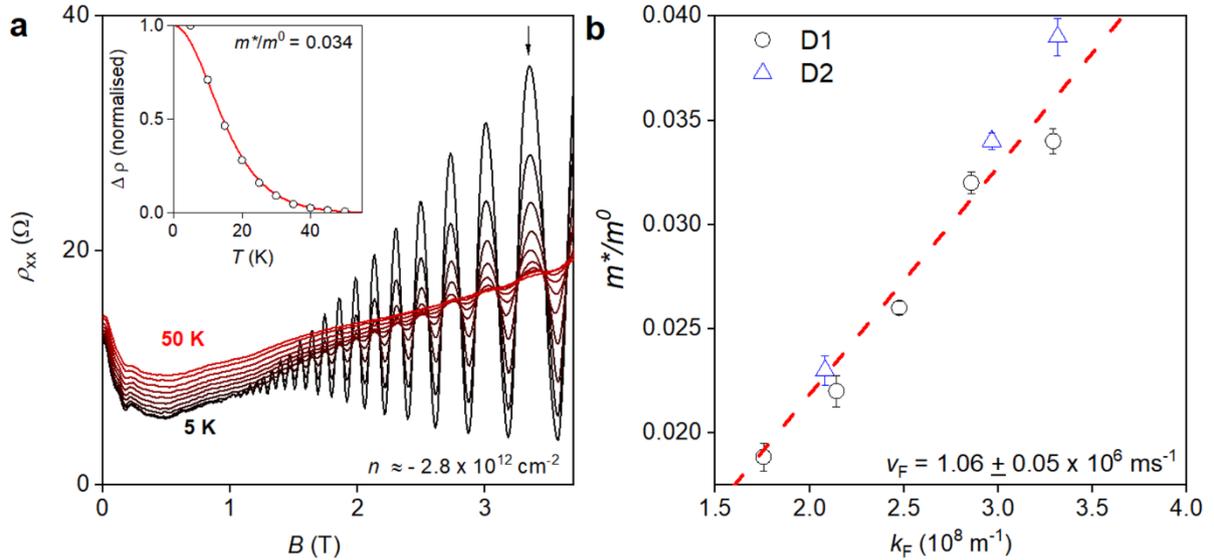

**Figure S4| Temperature dependence of Shubnikov de Hass oscillations. a,** Magnetoresistance $\rho_{xx}(B)$ measured at a fixed hole density $n$ = 2.8 x $10^{12}$ cm$^{-2}$ for different $T$ in 5 K steps. We subtract the smooth background and analyse the amplitude of SdH oscillations (indicated by black arrow). Inset: $T$ dependence of the normalised amplitude $\Delta\rho(T)$ for the corresponding peak that is indicated by the black arrow in **a**. Open circles are experimental data points and solid red line is the standard fitting with the Lifshitz-Kosevich function. $m^0$ represents the free electron mass **b,** measurements of $m^*/m^0$ for different $n$ of holes in two different graphene devices; the black circles represent data obtained in one of our widest devices $W$ = 15 μm (Fig. 1c of the main text). The error bars represent

the error in fitting the Lifshitz-Kosevich formula. Red dashed line is a linear fit to the equation $m^* = \hbar k_F/v_F$ with a constant $v_F = 1.06 \times 10^6$ ms$^{-1}$.

## 6 Summary of quantum transport calculations

We use the Kubo approach[11] to determine the linear response of the oscillatory longitudinal magnetoresistivity, $\rho_{xx}$, of monolayer graphene due to the resonant absorption and emission of LA and TA acoustic phonons by the charge carriers in a magnetic field, **B** = (0,0,-B), applied perpendicular to the graphene sheet. The model corresponds to ohmic conditions with the carriers in thermal equilibrium with the lattice vibrations. The electronic spectrum becomes quantised into a series of unevenly spaced Landau levels (LLs) with index $N$ given by Eq. (S3). It is convenient to use the Landau gauge where **A** = (0, -Bx, 0). The carrier wave function in the $K^+$ valley and the conduction band is then given by the pseudospinor

$$\psi_N^{K^+} = \frac{1}{\sqrt{2}} \begin{pmatrix} \phi_{|N|}(x-X) \\ -\text{sgn}(N) i \phi_{|N|-1}(x-X) \end{pmatrix}, \quad (S7)$$

where $\phi$ are simple harmonic oscillator states along $x$ and plane waves along $y$ given by

$$\phi_N(x) = A_N H_N\left(\frac{x}{l_B}\right) \exp\left(\frac{x^2}{2l_B^2}\right) \exp(ik_y y), \quad (S8)$$

Here $A_N = 1/\sqrt{L_y l_B 2^N N! \sqrt{\pi}}$ is a normalisation constant and $H_N$ are the Hermite polynomials[12,9,13]. A similar relation applies for the valence band and the $K^-$ valley. In the Kubo approach, the contribution of the TA and LA acoustic phonon scattering to the magnetoconductance $\sigma_{xx}$ is given by

$$\Delta\sigma_{xx}^{t,l} = \frac{g_v g_s \pi e^2}{S^2 k_B T \hbar} \sum_{\mathbf{q}} \left(l_B^2 q_y\right)^2 |C(q)|^2 N_q(N_q + 1) \sum_{N,N'} \sum_{k_y, k_y'} [f(E_N - \hbar\omega_q^{t,l}) - f(E_N)] \delta(E_N - \hbar\omega_q^{t,l} - E_{N'}) \left|I_{N,N'}^{t,l}(k_y, k_y', \mathbf{q})\right|^2. \quad (S9)$$

Here the subscripts/superscripts $t$ and $l$ refer to contributions corresponding to the TA or LA phonons respectively, $g_v$ = 2 and $g_s = 2$ are the valley and spin degeneracies, $S = L_x L_y$ is the area of the device, $k_B$ is the Boltzmann constant, $T$ is the lattice temperature, $|C_{t,l}(q)|^2 = \hbar q/(2\rho v_{t,l})$ are the Fourier components of the scattering potential, $\rho = 7.6 \times 10^{-8}$ g cm$^{-2}$ is the mass density of graphene, $N_q = (\exp(\hbar\omega_q^{t,l}/k_B T) - 1)^{-1}$ is the Bose-Einstein distribution function for the phonons and $f(E) = (\exp((E-\mu)/k_b T) + 1)^{-1}$ is the Fermi-Dirac distribution of the carriers with chemical potential, $\mu$. The matrix elements $I_{N,N'}^{t,l}(\mathbf{q})$, are given by

$$I_{N,N'}^{t,l}(k_y, k_y', \mathbf{q}) = \int dS \psi_{k_y', N'}^* V_\mathbf{q}^{t,l} \psi_{k_y, N}.$$

(S10)

$V_\mathbf{q}^{t,l}$ describes the charge-carrier phonon coupling for the TA and LA phonons respectively[14,15,16,17] and has the form

$$V_{\mathbf{q}}^{t} = e^{i\mathbf{q}\cdot\mathbf{r}} \begin{pmatrix} 0 & -g_g e^{i2\varphi} \\ g_g e^{-i2\varphi} & 0 \end{pmatrix} \tag{S11}$$

and

$$V_{\mathbf{q}}^{l} = i e^{i\mathbf{q}\cdot\mathbf{r}} \begin{pmatrix} g_d(\mathbf{q}) & g_g e^{i2\varphi} \\ g_g e^{-i2\varphi} & g_d(\mathbf{q}) \end{pmatrix}, \tag{S12}$$

where $\varphi$ is the angle between the phonon wave-vector and the x axis, which in our model is defined to be along the zigzag edge of the graphene layer. Here, $g_g$ and $g_d(\mathbf{q})$ are the carrier-phonon coupling matrix elements corresponding to the gauge and deformation distortions of the graphene lattice. The off-diagonal matrix elements involving $g_g$ arise from pure shear-like distortions of the lattice which give rise to a "synthetic" gauge field in the Dirac equation. This is unaffected by screening and has been estimated using density functional theory to have a value in the range $g_g$ = 1.5 − 4.5 eV[14,15,16,17,18]. We obtain a good fit to the data with $g_g$ = 4 eV. The on-diagonal terms correspond to strain-induced distortions of the unit cells that change their areas, resulting in local redistributions in the carrier density, *n,* which screen the deformation potential terms $g_d(\mathbf{q})$. Therefore we write

$$g_d(q) = \frac{\tilde{g}_d}{\varepsilon(q)}, \tag{S13}$$

where $\tilde{g}_d$ is the unscreened "bare" deformation potential coefficient and $\varepsilon(q)$ is the phonon wave vector-dependent dielectric function. We use the Thomas-Fermi approximation for $\varepsilon(q)$ which gives

$$\varepsilon(q) = \varepsilon_r \left(1 + \frac{q_{tf}}{q}\right), \tag{S14}$$

where $q_{tf} = 4e^2 \sqrt{n_i \pi}/(4\pi \hbar v_F \varepsilon_r \varepsilon_0)$ is the inverse Thomas-Fermi screening radius. This takes into account screening by the dielectric environment of the graphene layer, with the dielectric constant $\varepsilon_r$ and by the free carriers in the graphene layer[18,19]. For free-standing graphene and $\varepsilon_r$ = 1, $q_{tf} \sim 8k_F$. For the case of magnetophonon resonance at high LL index, $q \sim 2k_F$ and $\varepsilon(q) \sim 5$ and thus the deformation potential is strongly suppressed[18]. In our experiments, the graphene layer is fully encapsulated by hBN and therefore we set $\varepsilon_r$ = 3.5 so that $\varepsilon(q) \sim 7.5$ and $g_d(q)$ is reduced further. However, we find that the resistivity is not very sensitive to $\varepsilon_r$ since the deformation potential is very effectively screened by the carriers in the graphene layer. We set $\tilde{g}_d$ to be 25 eV. We find that in the range from $\tilde{g}_d = 20 - 30$ eV the resistivity is not sensitive to this parameter due to the strong screening effect.

To model LL broadening, we replace the delta function in Eq. S9 by

$$\delta(E) \rightarrow \frac{1}{\Gamma\sqrt{2\pi}} \exp\left(-\frac{E^2}{2\Gamma^2}\right), \tag{S15}$$

where $E = E_N - E_{N'} - q\hbar v_{t,l}$. We use a Gaussian function to aid convergence of our calculation at high LL indicies. In the case of short-range scattering, for example by charged impurities, the broadening of the LLs depends on the square root of magnetic field[12,20,21,22]. Therefore, we set the broadening parameter $\Gamma = \gamma\sqrt{B}$. We obtain a good fit to the data with γ = 0.5 meV T$^{-1/2}$ (see Fig. 4b of the main text).

In the Hall regime, the longitudinal magnetoresistivity is given by $\rho_{yy} = \sigma_{xx}/(\sigma_{xx}\sigma_{yy} + \sigma_{xy}^2)$, where the component $\sigma_{xy} = ne/B$. Since $\sigma_{xx} = \sigma_{yy} \ll \sigma_{xy}$ the oscillatory part of $\rho_{xx}$ due to TA and LA phonon scattering is given by

$$\Delta\rho_{xx} = \left(\frac{B}{ne}\right)^2 (\sigma_{xx}^l + \sigma_{xx}^t) \tag{S16}$$

to a good approximation. Figure S5 shows $\Delta\rho_{xx}$ calculated for three representative carrier densities, 6, 7.5 and 9 x $10^{12}$ cm$^{-2}$. We find that $\Delta\rho_{xx}(B)$ has a form and amplitude which agrees well with oscillations observed in the measurements shown in the main text. The magnetic field values of the position of the peaks correspond closely to the classical resonance condition in Eq. 1 of the main text and Eq. S5; The slight difference between the classical and quantum models arises from the form of overlap integral between the corresponding Landau-quantised wavefunctions. The peaks are periodic in 1/B with a frequency, $B_F$, that has a linear dependence on $n$ (S6) The plot shows that the contribution from the LA phonons to the total resistivity is relatively small and appears only as the $p$ = 1 peak in $\Delta\rho_{xx}$ (labelled by the blue arrow in Fig. S5). This is due to two main factors: first, the suppression of the deformation part of the electron-phonon coupling matrix due to free carrier screening (see Eq. S12); second, the energy of the LA phonon is larger than that of the TA phonon when the condition for magnetophonon resonance is satisfied. Therefore, at a given temperature, there is a lower population of LA phonons than TA phonons for carriers to absorb and similarly there are fewer carriers at high enough energy to emit an LA phonon compared to those that can emit a TA phonon. If the deformation part of the electron-phonon were not screened, $\rho_{xx}(B)$ would be dominated by the contribution from the LA phonons, highlighting the importance of the strong screening of LA phonon scattering in graphene.

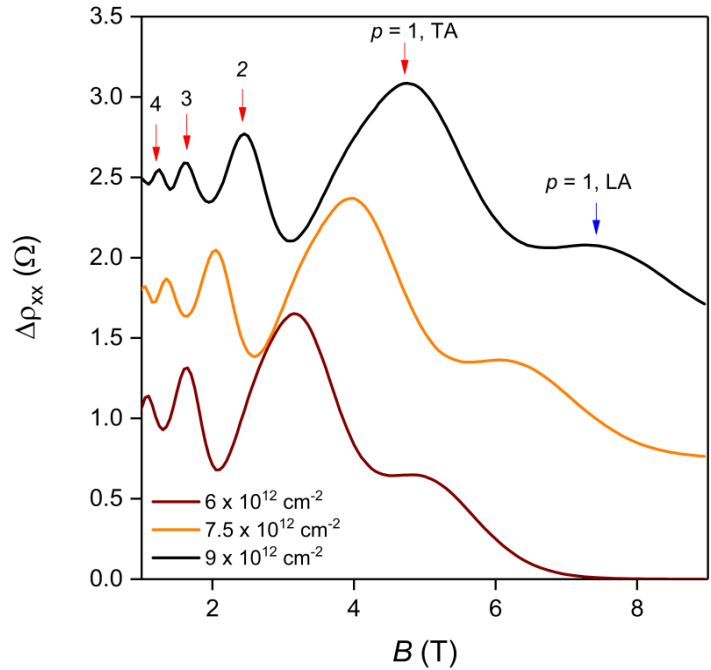

**Figure S5|** Calculated $\Delta\rho_{xx}(B)$ for a three different $n$ and $T$ = 70 K, using $v_t$ = 13.6 kms$^{-1}$ and $v_l$ = 21.4 kms$^{-1}$ taken from ref. 17 and the parameters specified in the text. The integers $p$ correspond to resonant inter-LL scattering around the Fermi energy, with $p = |N - N'|$. The curves are offset for clarity.